\documentclass[twocolumn, prb]{revtex4}
 \newif\ifGALLEYversion\GALLEYversionfalse

\usepackage{ifthen}
\usepackage{graphicx}
\usepackage{amsmath}

\usepackage{color}

\ifthenelse{\boolean{GALLEYversion}}{
    \marginparwidth 2.7in
    \marginparsep 0.5in
    \def\abm#1{\marginpar{\small AB: #1}}
    \def\thm#1{\marginpar{\small HT: #1}}
    \def\rdm#1{\marginpar{\small RD: #1}}
    \def\akm#1{\marginpar{\small AK: #1}}
}{
    \def\abm#1{\relax}
    \def\thm#1{\relax}
    \def\rdm#1{\relax}
    \def\akm#1{\relax}
}

\usepackage{graphicx}

\begin{document}

\title{Possible routes for synthesis of new boron-rich Fe-B and Fe$_{1-x}$Cr$_x$B$_4$ compounds}

\author{A.F. Bialon,$^{1}$ T. Hammerschmidt,$^{1}$ R. Drautz,$^{1}$ S. Shah,$^{2}$, E.R. Margine,$^{2}$ and A.N. Kolmogorov$^{2}$}

\affiliation{$^{1}$Atomistic Modelling and Simulation, ICAMS, Ruhr-Universit\"at Bochum, D-44801 Bochum, Germany}
\affiliation{$^{2}$Department of Materials, University of Oxford, Parks Road, Oxford OX1 3PH, United Kingdom}

\date{\today}

\begin{abstract}
  {We use {\it ab initio} calculations to examine thermodynamic
  factors that could promote the formation of recently proposed unique
  oP10-FeB$_4$ and oP12-FeB$_2$ compounds. We demonstrate that these
  compact boron-rich phases are stabilized further under pressure. We
  also show that chromium tetraboride is more stable in the new oP10
  rather than the reported oI10 structure which opens up the
  possibility of realizing an oP10-(Fe$_x$Cr$_{1-x})$B$_4$
  pseudobinary material. In addition to exhibiting remarkable
  electronic features, oP10-FeB$_4$ and oP12-FeB$_2$ are expected to
  be harder than the known Fe-B compounds commonly used for hard
  coating applications.}
\end{abstract}

\maketitle



The renewed interest in transition metal (TM) borides stems, in part,
from the materials' potential to serve as hard, wear-resistant,
chemically inert coatings.\cite{ReB2, Correlation, SHvsUIC, OsRuB2,
Advances} Combination of the TM and boron ensures a high valence
electron density and a pronounced covalent bonding resulting in the
compounds' exceptional hardness; for example, ReB$_2$ has been
recently demonstrated to be the first metal-based bulk material that
can scratch diamond.\cite{ReB2,Advances} Metallicity and covalency are
also key ingredients for phonon-mediated superconductivity and TM
borides have received a lot of attention\cite{MoB4} following the
discovery of a remarkable MgB$_2$ superconductor.\cite{MgB2_exp}

Iron borides have two particularly important industrial
applications. First, homogeneously dispersed second-phase particles
of Fe$_2$B are known to improve the tensile strength of low-carbon
steels\cite{ultrahighB} while pseudobinary Fe$_3$(B,C) and
Fe$_{23}$(B,C)$_6$ \cite{prec1,prec2} precipitates harden high-carbon
steels.\cite{prec_hardenability} Second, Fe-B-based hard protective
coatings are produced directly on the surface of steel via the process
of boriding.\cite{Coatings1, Coatings2, Coatings3} During this
thermochemical process boron diffuses into the steel forming either a
single-phase (Fe$_2$B) or a duplex-phase (Fe$_2$B+FeB) coating
layer.\cite{Coatings2} The two FeB and Fe$_2$B compounds have been
shown to crystallize in the oP8 (or the related
oS8)\cite{CrB_pure_and_CrB_inter} and tI12 configurations,
respectively, and are the only low-temperature ground states listed in
the latest experimental phase diagram.\cite{FeB-PhD} Synthesis of new
boron-rich Fe-B compounds could have technological implications as the
hardness of metal borides tends to increase with boron content. So
far, except for the observation of a metastable FeB$_{49}$
intercalation compound\cite{FeB49} only two studies reported on the
synthesis of amorphous\cite{aFeB2} and the AlB$_2$-type \cite{FeB2}
iron diboride but these compounds have not been reproduced.

Application of advanced compound prediction methods has recently
allowed us to identify oP12-FeB$_2$ and oP10-FeB$_4$ candidate ground
states with unique crystal structures (Fig. 1b,c) that are stable relative to
the known compounds.\cite{PRL} oP12-FeB$_2$ was predicted to be the
first metal diboride semiconductor while oP10-FeB$_4$ was shown to
have the necessary features to exhibit phonon-mediated
superconductivity with a $T_c$ of 15-20~K. The aim of this study is to
examine synthesis routes that could lead to the discovery of the new
materials. We find that both compounds are stabilized further under
pressure while the oP10 Fe-based phase could be realized in the
(Fe$_x$Cr$_{1-x}$)B$_4$ pseudobinary form. We also observe that
calculated elastic constants of the predicted Fe-B compounds are
higher than those in the known FeB and Fe$_2$B materials.

We use the projector augmented wave method\cite{PAW} as implemented in
{\small VASP}\cite{VASP} and carry out full structural and spin
relaxation for all compounds; relevant $M$B$_2$ and $M$B$_4$ metal
borides show no magnetic ordering and their phonon and electron-phonon
($e$-ph) properties are examined without spin polarization. The chosen
energy cutoff of 500~eV and dense Monkhorst-Pack
$k$-meshes\cite{MONKHORST_PACK} ensure numerical convergence of
formation energy differences to typically 1-2 meV/atom. We employ the
Perdew-Burke-Ernzerhof (PBE) exchange-correlation (xc)
functional\cite{PBE} within the generalized gradient approximation
(GGA). Vibrational corrections to Gibbs energy are calculated with
PHON.\cite{PHON} The strength of the $e$-ph coupling for oP10-CrB$_4$
is evaluated within the linear response theory using the {\small
Quantum-ESPRESSO} package.\cite{PWscf} The boron ground state is
simulated as $\alpha$-B, known to be stable at medium pressures up to
$\sim19$ GPa.\cite{aB,gB} The iron ground state is taken as bcc and
hcp at 0 and 20 GPa, respectively.\cite{FePressure}
\begin{figure}[t]
\begin{center}
\hspace{-1.0 cm}
\includegraphics[width=75 mm,angle=0]{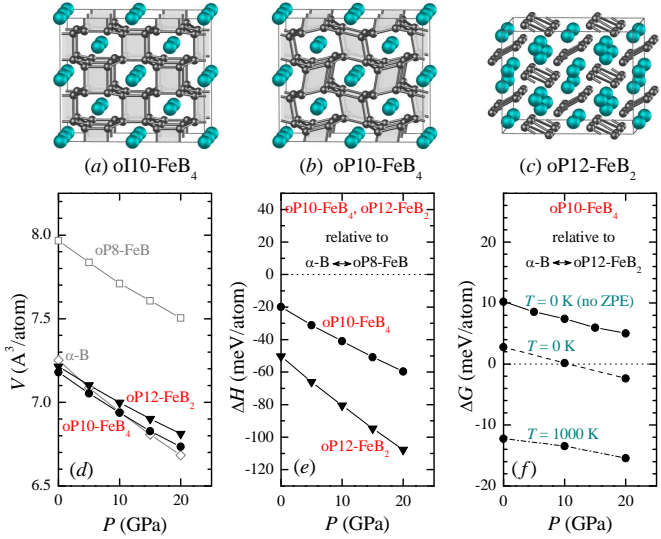}
\caption{ \small (Color online) $a$--$c$) Boron-rich structures: iron
and boron are shown as large cyan and small black spheres,
respectively. $d$) Atomic volume as a function of pressure. $e$)
Relative enthalpy of oP12-FeB$_2$ and oP10-FeB$_4$ candidate phases
w.r.t. the $\alpha$-B$\leftrightarrow$oP8-FeB tie-line. $f$) Relative
Gibbs energy (with vibrational contributions included using
PHON\cite{PHON}) of oP10-FeB$_4$ w.r.t.
$\alpha$-B$\leftrightarrow$oP12-FeB$_2$ at 0 K with and without zero
point energy (ZPE) and at 1000 K.}
\label{fig1}
\end{center}
\end{figure}

The peculiar oI10-TMB$_4$ phases comprised of tetragonal B nets
(Fig. 1a) were previously examined within the extended H\"{u}ckel
method; it was concluded that maximum binding in the $3d$ series is
achieved for Cr and that the electron-rich Fe, Co, and Ni tetraborides
may be unstable in this configuration.\cite{TMB4} Our calculations
showed \cite{PRL} dynamical instability of the oI10-FeB$_4$ phase
which gains a considerable 0.13 eV/f.u. in enthalpy by transforming
into the oP10 structure (Fig. 2b). In the present study we observe a
similar behaviour in the Cr-B system: oI10-CrB$_4$ is found to be both
dynamically (Fig. 1 in Ref. \onlinecite{SuppMat}) and
thermodynamically (by 0.03 eV/f.u.) unstable relative to oP10.
Although the boron network undergoes a significant distortion in the
oI10$\rightarrow$oP10 transformation, the lattice parameters and the
simulated powder diffraction patterns (Fig. 2 in
Ref. \onlinecite{SuppMat}) remain close which may explain why the
structure of CrB$_4$ was originally solved as oI10. These findings
suggest that Fe$_x$Cr$_{1-x}$B$_4$ compositions may assume the oP10
structure as well under standard synthesis conditions. Indeed, there
were no special requirements regarding the heat treatment or the
starting materials (apart from cold compacting of the $M$B powders
\cite{car-powder-fe-cr,car-powder-feb-crb}) for the past synthesis of
ternary Fe-Cr-B materials, e.g., the ordered metal-rich Mn$_4$B-type
alloy\cite{beerntsen-fecrb-mn4b} or the oP8-FeB- and oS8-CrB-type
compounds with partial substitutions of the host
metal.\cite{car-powder-feb-crb,car-powder-fe-cr}

The effect of the composition on the thermodynamic and electronic
properties of the ordered oP10-Fe$_x$Cr$_{1-x}$B$_4$ compounds is
investigated via supercell simulations,\cite{MnB4} see Fig. 2. We
observe a sizeable stabilization, up to 28 meV/f.u., of the oP10
structure at $x=0.5$, Fig. 2a. The configurational entropy
contribution from the disorder in population of metal sites is
comparable at elevated temperatures: $\Delta G_{\text
conf}(T)=-k_BT[x\ln x+(1-x)\ln(1-x)]$ would be 60 meV/f.u. at $T$=1000
K and $x=0.5$ if all decorations were degenerate in energy. Hence, the
resulting ordering of Fe and Cr on the metal sublattice will depend on
the quenching conditions. The density of states (DOS) is found to be
sensitive to metal site population patterns at the Fe-rich end and
drops rapidly in going from pure FeB$_4$ (1.0 states/(spin eV atom))
to pure CrB$_4$ (0.18 states/(spin eV atom)), Fig. 2b. Our linear
response theory calculations\cite{PWscf,kq_meshes} give a small $e$-ph
coupling ($\lambda\approx 0.15$) and a negligible critical temperature
($T_c\ll 1$ K) in oP10-CrB$_4$ indicating that superconductivity in
this pseudobinary will require high Fe concentrations.
\begin{figure}[t]
\begin{center}
\includegraphics[width=88mm,angle=0]{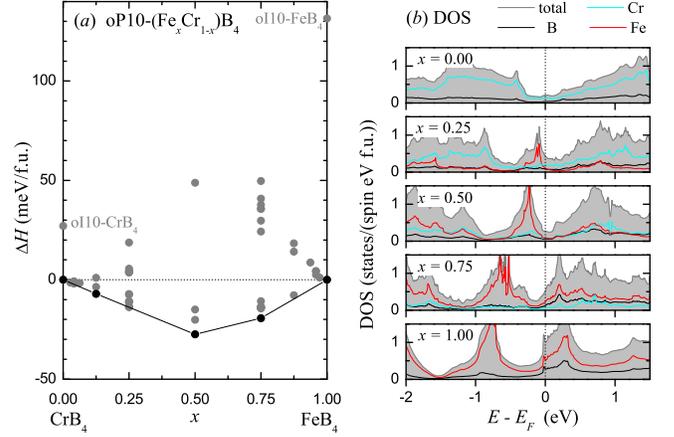}
\caption{ \small (Color online) $a$) Relative stability of the
pseudobinary oP10-Fe$_x$Cr$_{1-x}$B$_4$ compound. $b$) Total and
projected density of states (DOS) for lowest enthalpy compounds.}
\label{fig2}
\end{center}
\end{figure}

Application of medium pressures (a few GPa) in multi-anvil or diamond
anvil cell setups may promote the formation of the materials in the
predicted configurations by improving the compounds' thermodynamic
stability or the reaction kinetics. Successful examples of this
synthesis route include boron-rich CaB$_4$ and NdB$_6$ or metal-rich
Fe$_2$B materials.\cite{liu-cab4,zhao-ndb6,yao-fe3b-fe2b} We have examined the
response to the hydrostatic pressure of over 40 known and proposed
ambient-pressure $M$-B structure types listed in
Ref. \onlinecite{SuppMat} by calculating their formation enthalpies at
$P=20$ GPa. We find that oP8-FeB is the lowest-enthalpy phase and that
the necessary condition for a boron-rich phase to be thermodynamically
stable is to lie below the oP8-FeB$\leftrightarrow$$\alpha$-B
tie-line. Due to the remarkable compactness (Fig. 1d) of the predicted
oP10-FeB$_4$ and oP12-FeB$_2$ phases (a respective 5\% and 7\%
reduction in atomic volume compared to mixture of $\alpha$-B and FeB)
pressure does lead to much lower relative formation enthalpies for
both compounds (Fig. 1e). Above $P=10$ GPa, oP10-FeB$_4$ becomes
thermodynamically stable relative to $\alpha$-B and oP12-FeB$_2$ at
{\it all} temperatures(Fig. 1f).

\begin{table}[b]
\begin{center}
\begin{tabular}[t]{cp{1mm}ccccccccccc}\hline\hline
            && Fe$_2$B && \multicolumn{3}{c}{FeB} && FeB$_2$ && FeB$_4$ && ReB$_2$ \\
            &&  tI12   &&  oP8  &  oS8 & tI16 && oP12 && oP10 &&  hP6 \\\hline
 $B$(GPa)   &&  228    &&  287  &  257 &  286 &&  311 &&  274 &&  339 \\
 $G$(GPa)   &&  143    &&  138  &  153 &  157 &&  231 &&  187 &&  266 \\
 $E$(GPa)   &&  355    &&  358  &  383 &  399 &&  556 &&  457 &&  632 \\
  $\nu$     && 0.24    && 0.29  & 0.25 & 0.27 && 0.20 && 0.22 && 0.19 \\\hline
\end{tabular}
\caption{\small Calculated bulk ($B$), shear ($G$), and Young ($E$)
moduli and Poisson ratio ($\nu$) for known and predicted Fe-B
phases. hP6-ReB$_2$ is added for comparison.}
\label{table_elacon}
\end{center}
\end{table}
The compactness of these Fe-B phases also opens the
perspective of obtaining new metal-based hard materials. It has been
argued that the extraordinary hardness of the ReB$_2$ material arises
from an efficient packing of B into hcp-Re that results in only a 5\%
expansion of the metal lattice and, consequently, in the shortest
TM-TM distances for any TM diboride.\cite{ReB2} The importance of
having strong TM-TM and TM-B bonds is that the (0001) plane in
hP6-ReB$_2$ supports the weakest stress. In the predicted
oP12-FeB$_2$, the TM-TM (TM-B) bonds are shorter by 11\% (9\%) not
only because of the smaller size of Fe but also because of the
break-up of the B layers into B chains which tilt to accommodate the
metal atoms (Fig. 3 in Ref. \onlinecite{SuppMat}). This leads to a more
pronounced relative deviation from Vegard's law\cite{Vegard} in
oP12-FeB$_2$ (-16.4\%) compared to that in hP6-ReB$_2$ (-6.4\%). A
qualitative analysis of the materials hardness can be done by
comparing shear moduli, as they have been shown to correlate with
hardness in some cases.\cite{Correlation} The shear moduli in the
predicted B-rich phases are at least as high as in the known oP8-FeB
and tI12-Fe$_2$B compounds (Table I). Although elastic constants alone
are not sufficient for a quantitative prediction of materials
hardness,\cite{Lazar} it is encouraging to find the elastic moduli and
the Poisson ratio in oP12-FeB$_2$ to be comparable to those in
hP6-ReB$_2$ (see Tables I and II in Ref. \onlinecite{SuppMat} for
comparison of experimental and theoretical values).  

In summary, our high-throughput simulations spanning a large library
of known and proposed structures predict consistently that new compounds
should form under accessible synthesis conditions in such a common
binary system as Fe-B. Ground state search under ambient or elevated
pressures can be expanded further by performing unrestricted
structural optimization driven, e.g., by evolutionary algorithms
\cite{PRL} for larger unit cells and other Fe-B compositions. Due to
the known exceptional complexity of metal boride
structures\cite{ak09} experimental input may be key for
determination of the true ground states.

ANK and SS acknowledge the support of the EPSRC. AFB, TH and RD
acknowledge financial support through ThyssenKrupp AG, Bayer
MaterialScience AG, Salzgitter Mannesmann Forschung GmbH, Robert Bosch
GmbH, Benteler Stahl/Rohr GmbH, Bayer Technology Services GmbH and the
state of North-Rhine Westphalia as well as the EU in the framework of
the ERDF.



\begin{thebibliography}{99}
\bibitem{MnB4} 
Since W$_{0.5}$Os$_{0.5}$B$_2$ is known to assume the ReB$_2$
prototype\cite{SHvsUIC} we also examined ordered
Fe$_{0.5}$Cr$_{0.5}$B$_4$ in the related mS10-MnB$_4$ supercell
configurations but found them to be less stable by 22 meV/f.u
w.r.t. the CrB$_4$-FeB$_4$ tie-line.
\bibitem{Vegard}
The relative deviation is defined as $V_{M{\text B}_2}/(2/3V_{\text
B}+1/3V_M)-1$; all volumes are per atom.
\end{thebibliography}
\end{document}


\title{Supplementary material to: Possible routes for synthesis of new boron-rich Fe-B and Fe$_{1-x}$Cr$_x$B$_4$ compounds}

\author{A.F. Bialon,$^{1}$ T. Hammerschmidt,$^{1}$ R. Drautz,$^{1}$ S. Shah,$^{2}$, E.R. Margine,$^{2}$ and A.N. Kolmogorov$^{2}$}

\affiliation{$^{1}$Atomistic Modelling and Simulation, ICAMS, Ruhr-Universit\"at Bochum, D-44801 Bochum, Germany}
\affiliation{$^{2}$Department of Materials, University of Oxford, Parks Road, Oxford OX1 3PH, United Kingdom}

\date{\today}

\maketitle




\begin{figure}[!t]
\begin{center}
\includegraphics[width=80 mm,angle=0]{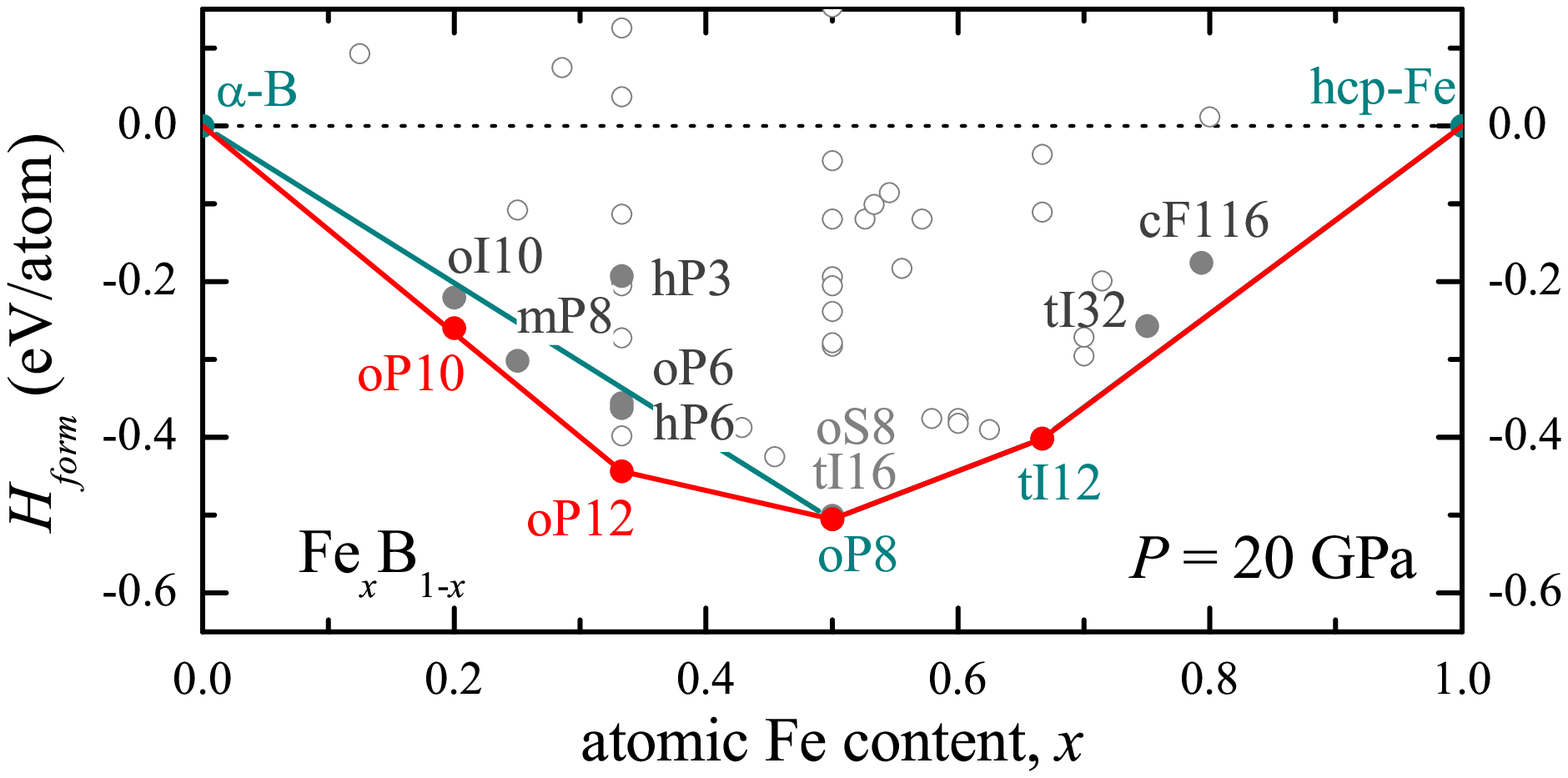}
\caption{ \small (Color online) Calculated formation enthalphy of Fe-B
compounds at $P=20$ GPa. The cyan and red labels correspond to known
and proposed ground states, respectively.}
\label{sup1}
\end{center}
\end{figure}
\begin{figure}[!t]
\begin{center}
\includegraphics[width=80 mm,angle=0]{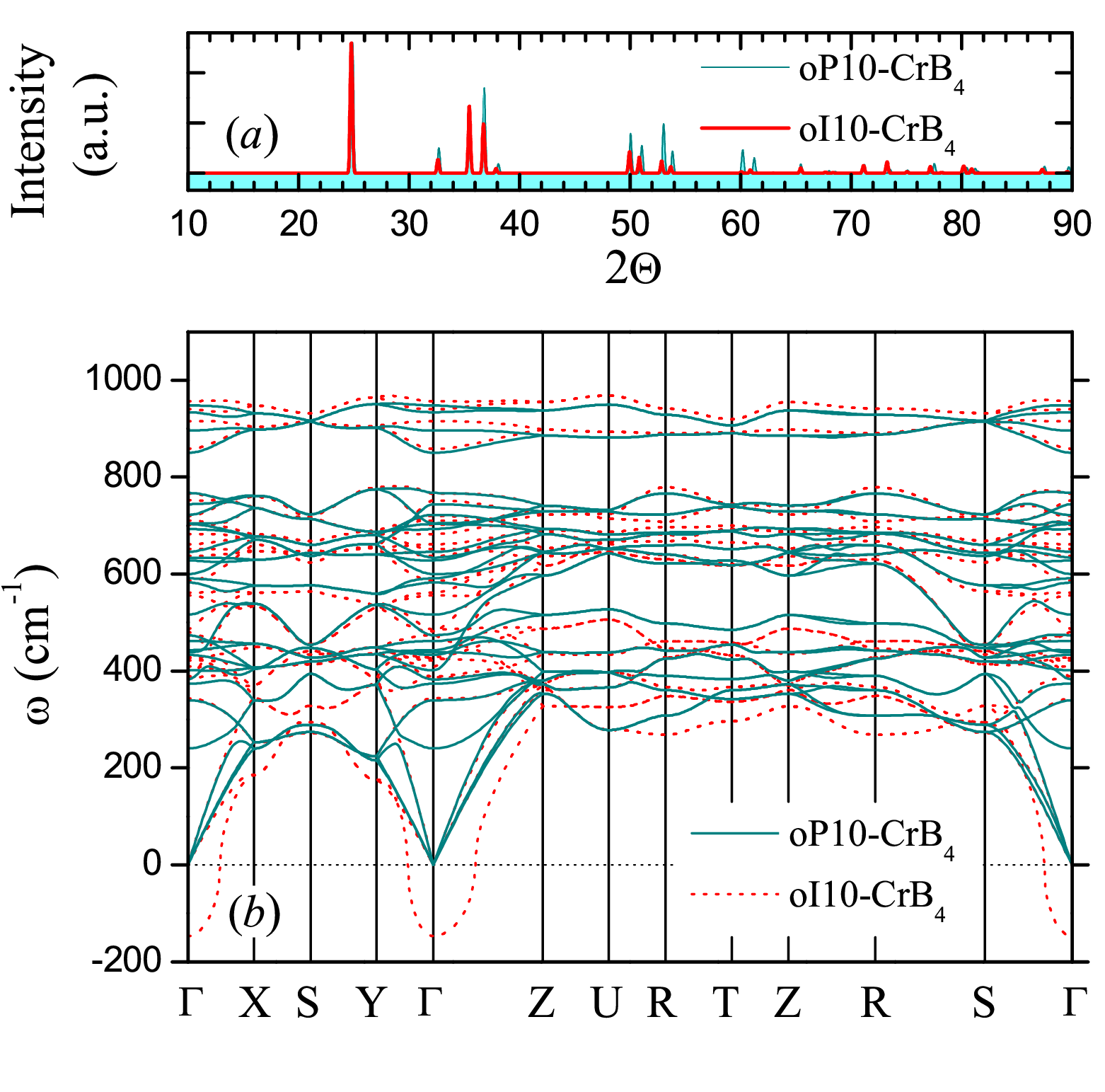}
\caption{ \small (Color online) Comparison of competing oI10- and
oP10-CrB$_4$ structures with relaxed lattice vectors of $a=4.749$ \AA,
$b=5.488$ \AA, $c=2.852$ \AA\ and $a=4.723$ \AA, $b=5.474$ \AA,
$c=2.851$ \AA, respectively: $a)$ simulated x-ray patterns for
$\lambda=1.5418$ \AA; $b)$ calculated phonon dispersons showing
dynamic instability of oI10-CrB$_4$ at $q=\Gamma$.}
\label{sup2}
\end{center}
\end{figure}

{\it Library of structure types.} We have considered over 40 commonly
seen $M$-B and $M$-C structure types listed in the Inorganic Crystal
Structure Database (ICSD).\cite{ICSD} The list below is ordered by
composition and specifies the Pearson symbol and the prototype for
each entry. 1:12: cF52-UB$_{12}$; 1:7: oI64-MgB$_7$; 3:20:
oS46-Na$_3$B$_{20}$; 1:6: cP7-CaB$_6$; 3:14: tI160-Li$_3$B$_{14}$;
1:4: tP20-UB$_4$, oI10-CrB$_4$, mS10-MnB$_4$; 1:3: tP20-LiB$_3$,
hP16-Mo$_{0.8}$B$_3$; 2:5: hR21-Mo$_2$B$_{4.65}$; 1:2: hP3-AlB$_2$,
oP6-RuB$_2$, hR18-MoB$_2$, oP12-PbCl$_2$, hP6-ReB$_2$, hP6-MoS$_2$;
2:3: oS20-V$_2$B$_3$, hP10-Ru$_2$B$_3$; 3:4: oI14-Ta$_3$B$_4$; 5:6:
oS22-V$_5$B$_6$; 1:1: oP8-FeB-b, oS8-PtB$_{0.67}$, tI16-MoB, oS8-TlI,
oP8-TiSi, hP2-WC; 5:4: hP18-Rh$_5$B$_4$; 11:8: oP38-Ru$_{11}$B$_{8}$;
3:2: tP10-U$_3$Si$_2$, aP30-Rh$_9$B$_{5.5}$; 5:3: tI32-Cr$_5$B$_3$;
2:1: tI12-CuAl$_2$; 7:3: hP20-Th$_7$Fe$_3$, oP40-Mn$_7$C$_3$; 5:2:
mS28-Mn$_5$C$_2$; 3:1: oP16-Fe$_3$C, oS16-Re$_3$B, tI32-Ni$_3$P; 23:6:
cF116-Cr$_{23}$C$_6$; 4:1: tP10-Be$_4$B. The structures proposed in
our previous study\cite{PRL} include 1:4: oP10-FeB$_4$, 1:3:
mP8-FeB$_3$, 1:2: oP12-FeB$_2$.
\begin{figure}[!h]
\begin{center}
\includegraphics[width=75 mm,angle=0]{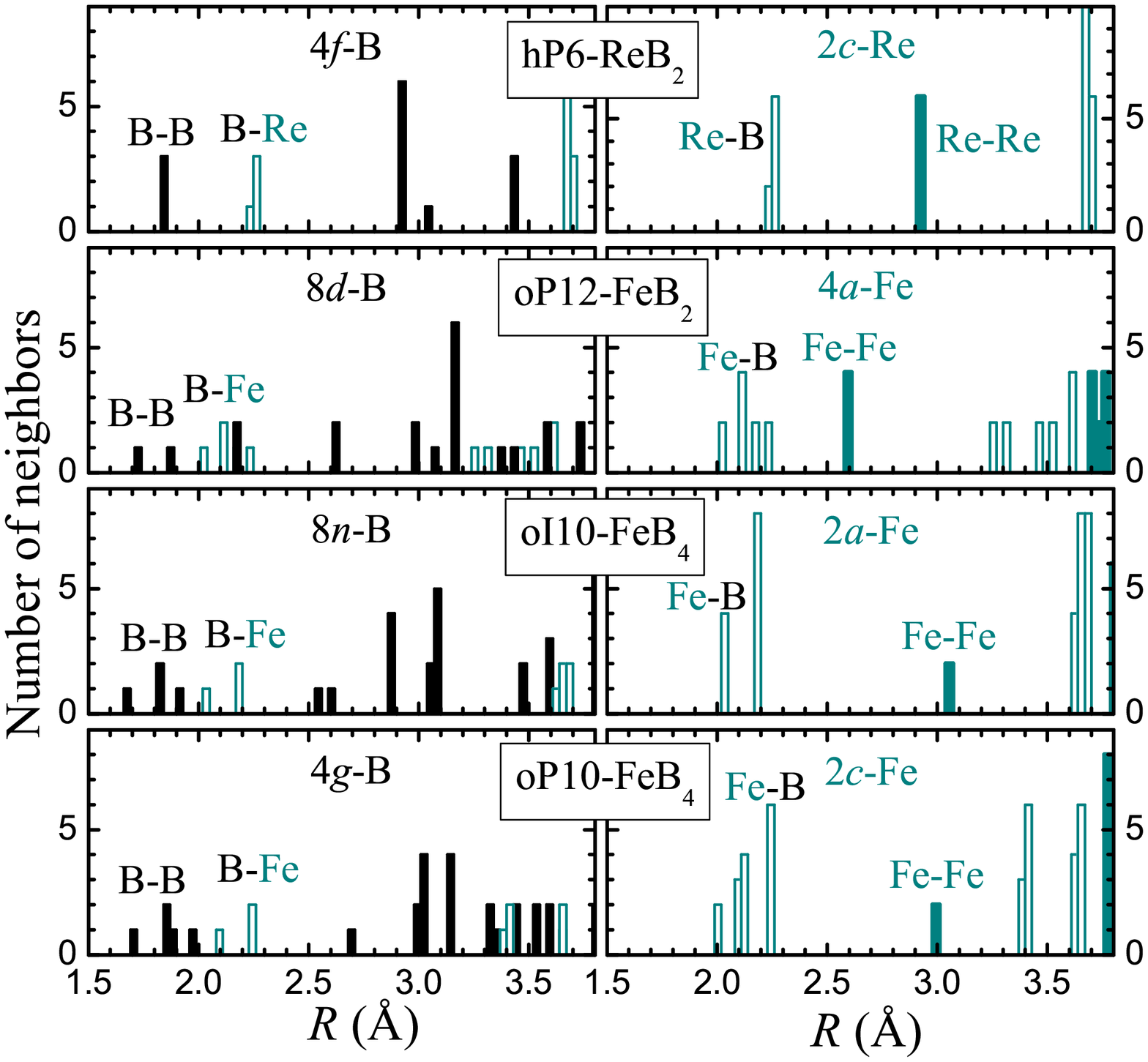}
\caption{ \small (Color online) Nearest neighbor histograms in fully
relaxed boron-rich Re-B and Fe-B structures. In oP10-FeB$_4$ the
neighbors are shown only for one of the two $4g$ boron sites.}
\label{sup3}
\end{center}
\end{figure}

{\it Comparison to published data}. We use available information on
stable iron and rhenium borides at ambient pressure to benchmark the
performance of the generalized gradient approximation\cite{PBE}
employed in this study. For tI12-Fe$_2$B and oP8-FeB our calculated
magnetic moments, 1.85 and 1.16 $\mu_B$/Fe, are in good agreement with
previously measured values of 1.91 and 1.03 $\mu_B$/Fe,
respectively.\cite{perkins_1974} As shown in Table \ref{hof-table},
for tI12-Fe$_2$B the calculated heats of formation are close to the
previously calculated values but are noticeably larger in magnitude
compared to the measured ones; for oP8-FeB we observe an excellent
agreement between theory and experiment. The elastic constants are
determined by fitting the total energy of distorted unit cells to
polynomials of second-order in strain. In order to determine all
independent elastic constants for a given cell type, the same number
of different strains is used. For each applied strain in the range
$[-0.03:0.03]$ with $0.01$ steps the internal coordinates were fully
relaxed. We find a generally good agreement between the present and
previously reported values for the elastic constants and moduli, see
Table \ref{ela-tab}.

\begin{table}[!t]
\begin{center}
\begin{tabular}[t]{ccccccc}\hline\hline
  && && $\Delta H$ (kJ/mol of atoms) && Ref. \\\hline
  tI12-Fe$_2$B && exp. && -22.30$^{a)}$ && \onlinecite{palumbo_2001} \\
  && exp. && -22.60$^{b)}$              && \onlinecite{palumbo_2001} \\
  && exp. && -23.71\phantom{$^{b)}$}    && \onlinecite{xiao_2009} \\
  && exp. && -(9.48-22.62)              && \onlinecite{gordienko_2002} \\
  && theo. && -32.09\phantom{$^{b)}$}   && \onlinecite{xiao_2009} \\
  && theo. && -29.67$^{c)}$             && \onlinecite{widom} \\
  && theo. && -30.30$^{c)}$             && This work \\
  &&&&&& \\
  oP8-FeB && exp. && -35.60$^{a)}$ && \onlinecite{palumbo_2001} \\
  && exp. && -32.30$^{b)}$ && \onlinecite{palumbo_2001} \\
  && theo. && -35.48$^{c)}$ && \onlinecite{widom} \\
  && theo. && -36.08$^{c)}$ && This work \\\hline\hline
\end{tabular}
\caption{Heat of formation for two stable compounds in the Fe-B system
  (\,$^{a)}$ reference states: $\alpha$-Fe, $\beta$-B, $T$=298\,K; $^{b)}$
  reference states: $\gamma$-Fe, $\beta$-B, $T$=1385\,K; $^{c)}$ values
  originally given in eV/atom).}
\label{hof-table}
\end{center}
\end{table}

\begin{widetext}
\begin{center}
  \begin{table*}[h!]
    \hfill{}
    \begin{tabular}[t]{cccccccccccccccccccc}\hline\hline
      & && & $C_{11}$ & $C_{12}$ & $C_{13}$ & $C_{22}$ & $C_{23}$ &
      $C_{33}$ & $C_{44}$ & $C_{55}$ & $C_{66}$ & & $B$ & $G$ & $E$ &
      $\nu$ & & Ref. \\\hline
      tI12-Fe$_2$B && theo. && 389.8 & 169.6 & 218.5 & & & 282.2 & 114.0 & &
      89.2 & & 249.7$^{a)}$ & 60.2$^{a)}$ & 184.4 & 0.388 & &
      \onlinecite{xiao_2010} \\
      && theo. & & & & & & & & & & & & 331.04$^{b)}$ & 152.77 & 397.22 & 0.3 & &
      \onlinecite{xiao_2009} \\  
      && theo. & & 413 & 154 & 132 & & & 389 & 148 & & 157 & & 228
      & 143 & 355 & 0.24 & & This work \\
      && exp. &&&&&&&&&&&& & & 343 &&& \onlinecite{ozbek_2002} \\
      && exp. &&&&&&&&&&&& & & 290 &&& \onlinecite{kuni_1971,golanski_1995} \\
      && exp. &&&&&&&&&&&& & & 300 &&& \onlinecite{lyak_1973} \\
      && exp. &&&&&&&&&&&& 164 & & &&& \onlinecite{chen_2004} \\
      &&&&&&&&&&&&&&&&&&& \\
      oP8-FeB && theo. & & 371 & 250 & 188 & 431 & 209 & 505 & 207 & 118 & 194
      & & 287 & 138 & 358 & 0.29 & & This work \\
      && exp. &&&&&&&&&&&& & & 284 &&& \onlinecite{ozbek_2002} \\
      && exp. &&&&&&&&&&&& & & 350 &&& \onlinecite{kuni_1971} \\
      && exp. &&&&&&&&&&&& & & 600 &&& \onlinecite{lyak_1973} \\ 
      && exp. &&&&&&&&&&&& & & 125-624 &&& \onlinecite{culha_2008} \\
      &&&&&&&&&&&&&&&&&&& \\
      hP6-ReB$_2$ && theo. & & 641 & 159 & 128 & & & 1037 & 271 & & &
      & 350$^{c)}$ & 283$^{c)}$ & & & &\onlinecite{zhou_reb2_2007} \\
      && theo. & & 631.3 & 158.4 & 133.8 & & & 1015.0 & 257.0 & & & &
      347.7$^{c)}$ & 273.5$^{c)}$ & & & &\onlinecite{zhang_reb2_2007} \\
      && theo. & & 689 & 170 & 141 & & & 1090 & 281 & & & & 369$^{a)}$
      & 294$^{a)}$ & 696 & 0.19 & &\onlinecite{zhao_reb2_2010} \\
      && theo. & & 632 & 154 & 131 & & & 1005 & 248 & & & & 339$^{a)}$ &
      266$^{a)}$ & 632 & 0.19 & & This work \\
      && exp. &&&&&&&&&&&& 360 & & &&& \onlinecite{chung_2007} \\
      && exp. &&&&&&&&&&&& 382 & 273 & &&& \onlinecite{levine_2009} \\
      && exp. &&&&&&&&&&&& 371$\pm$22 & 302$\pm$18 & 712$\pm$43 &&&
      \onlinecite{chung_2008}\\\hline\hline
    \end{tabular}
    \hfill{}
    \caption{Comparison of calculated elastic constants and moduli to
    previously reported theoretical and experimental values for
    selected known compounds. The elastic moduli were calculated using
    $^{a)}$ Voigt-Reuss-Hill theorem, $^{b)}$ equation of state, or
    $^{c)}$~Voigt theorem.}
    \label{ela-tab}
  \end{table*}    
\end{center}
\end{widetext}